\documentclass[a4paper,11pt]{article}
\usepackage{pos}
\usepackage{lmodern}
\usepackage{tikz}
\usepackage{tcolorbox}
\usepackage{amsmath}
\usepackage{slashed}
\usepackage{booktabs}

\title{BSM Developments and Tools}
\author*[a]{Tom\'as E. Gonzalo}

\affiliation[a]{Institute for Theoretical Particle Physics and Cosmology (TTK), RWTH Aachen University, D-52056 Aachen, Germany}

\emailAdd{gonzalo@physik.rwth-aachen.de}

\abstract{In this conference paper I introduce a selection of BSM tools and describe their most recent developments. I choose to focus on tools for the reinterpretation of LHC searches, tools that compute dark matter constraints, beyond-the-Standard-Model (BSM) inference tools, and tools that calculate amplitudes and cross sections directly from the Lagrangian.}

\FullConference{%
 
 The Ninth Annual Conference on Large Hadron Collider Physics - LHCP2021\\
 7-12 June 2021\\
 Online
}

\begin{document}

\begin{flushright}{TTK-21-40}\end{flushright}

\maketitle

\section{Introduction}

We are spoiled for choices when it comes to BSM tools. For many years the community has developed tools for every aspect of BSM physics, from various computations of observables, to abstract level tools able to handle the full Lagrangian of a theory. Due to the large amount of software available, in this conference article I will only describe a selection of tools and their recent developments. I will focus on tools for the recasting of LHC searches, dark matter (DM) tools, statistical inference tools and Lagrangian-level tools.

\section{Reinterpretation of LHC searches}

The reinterpretation of BSM searches has been a profilic field of study in particle physics. Significant advances have been made in the last few years in the interaction between the experimental and theoretical communities, and the exchange of information between them is at its highest, especially after the release of simplified and full likelihoods by CMS and ATLAS, respectively~\cite{LHCReinterpretationForum:2020xtr}. This has had profound positive effects on the whole community as well as the physics outcomes, as it has helped with the long-term preservation of the results and provides motivation for the design of new analyses. Consequently, many recasting tools have been developed, following two different approaches: the \textit{Simplified Models} approach, or the \textit{Fast Simulation} approach.

The \textit{Simplified Models} approach tackles the issue of recasting by decomposing a given model into simplified model topologies and comparing the predicted cross section times branching ratio with that reported by the experiments~\cite{Kraml:2013mwa}. As it does not require any simulation of collisions, this approach is extremely fast and perfectly suitable for large parameter fits. The prominent tool that implements this approach is \textsf{SModelS}, which is in fact responsible for pioneering the idea~\cite{Kraml:2013mwa}. \textsf{SModelS} currently contains the largest database of implemented analyses in the community, with over $100$ run I and II available analyses. With its latest version, \textsf{SModelS} can be interfaced with the full likelihoods provided by the ATLAS experiment~\cite{Alguero:2020grj}. Another notable tool that implements the Simplified Model approach is \textsf{DarkCAST}, a recasting tool for dark photon searches~\cite{Ilten:2018crw}.

On the other hand is the \textit{Fast Simulation} approach, which requires  to perform a Monte-Carlo simulation of the hard scattering process first, then drag the generated events through some form of detector modelling and lastly pass them through an event analysis framework. There are many available event analysis frameworks, some of them with native detector modelling. A non-exhaustive list of these include \textsf{MadAnalysis 5}~\cite{Conte:2012fm, Conte:2014zja, Dumont:2014tja, Conte:2018vmg, Araz:2019otb, Araz:2020lnp}, which can be used with a detector simulation or transfer functions and contains around 40 run I and II analyses; \textsf{CheckMATE}~\cite{Drees:2013wra,Dercks:2016npn}, which contains over 50 run I and II analyses, and was recently extended to support long-lived particle searches in addition to prompt particle searches; \textsf{ColliderBit}~\cite{GAMBIT:2017qxg}, which includes its own detector modelling in the form fast 4-vector smearing (\textsf{Buckfast}), and has a database of around 40 analyses, from runs I and II\footnote{\textsf{ColliderBit} is part of the \textsf{GAMBIT} framework, but it will soon be released as a standalone package}; and \textsf{Rivet}~\cite{Buckley:2010ar, Bierlich:2019rhm}, which has currently ``only'' 30 BSM analyses, but boasts a huge library of SM analyses, upwards of 800. In fact, it has been shown that SM inclusive measurements also set strong constraints on new physics, and the tool \textsf{Contur} has been developed for this purpose~\cite{Butterworth:2016sqg, Amrith:2018yfb, Buckley:2020wzk, Butterworth:2020vnb, Buckley:2021neu}. A comparison of the performance of these tools for a specific CMS search for supersymmetric (SUSY) particles~\cite{CMS:2018kag} was performed~\cite{Brooijmans:2020yij} and strong agreement was seen across all included tools.
  
\section{Dark Matter}

Searches for DM are a strong component of the physics agenda of the LHC experiments, but the strongest constraints on DM models arise often from  direct and indirect detection, as well as from the relic density (RD) of DM, and thus many tools have been developed over the years for that purpose\footnote{A thorough review of many DM tools can be found in \cite{Arina:2020alg}}. In recent times, there is a tendency towards the development of model-independent tools, able to constrain a large variety of BSM models. In addition, many existing and new tools attempt a comprehensive approach to constraints of DM, intending to simultaneously provide constraints from RD, direct and indirect detection. The most employed comprehensive tools developed by the community are \textsf{DarkSUSY 6}~\cite{Gondolo:2004sc, Bringmann:2018lay}, recently expanded to compute the RD beyond the standard kinetic equilibrium; \textsf{MicrOMEGAs}~\cite{Belanger:2001fz,Belanger:2004yn,Belanger:2006is,Belanger:2008sj,Belanger:2010gh,Belanger:2013oya,Belanger:2014vza}, whose most recent upgrade included computations of freeze-in RD; \textsf{MadDM}~\cite{Backovic:2013dpa, Backovic:2015cra, Ambrogi:2018jqj, Arina:2020kko}, part of the \textsf{MadGraph} toolchain; and \textsf{SuperISO Relic}~\cite{Arbey:2009gu, Arbey:2011zz, Arbey:2018msw}, which is widely used for SUSY models.

Even though comprehensive tools can cover all aspects of DM constraints, some more dedicated tools are better suited to compute a single constraint with higher accuracy. In the case of constraints from direct detection (DD), the front runner is \textsf{DDCalc}~\cite{GAMBITDarkMatterWorkgroup:2017fax, GAMBIT:2018eea}, which contains the largest collection of DD experiments to date, and can be used for complete as well as effective DM models, with RG evolution provided by \textsf{DirectDM}~\cite{Bishara:2017nnn, Brod:2017bsw}. In the indirect direction frontier the amount of available tools is too large to recount here. A short list of selected tools include \textsf{Capt'n General}~\cite{Kozar:2021iur}, which imposes constraints on DM capture in the Sun, \textsf{nulike}~\cite{IceCube:2012fvn, IceCube:2016yoy}, which calculates constraints from the DM annihilation to neutrinos, and \textsf{gamLike}~\cite{GAMBITDarkMatterWorkgroup:2017fax}, which deals with DM annihilation to $\gamma$-rays. Lastly one can impose constraints on DM models from cosmological observations. In particular, the energy injected into the Cosmic Microwave Background from DM annihilations in the early Universe can be computed with \textsf{DarkAges}~\cite{Stocker:2018avm}, part of the \textsf{ExoCLASS} branch of \textsf{CLASS}~\cite{Blas:2011aaa}.

\section{Inference and Sampling}

It has long been the goal of many particle theorists to assess the validity of BSM models in light of data. However, the vast quantities of data produced by the experiments, and the increasingly large parameter spaces of the models, makes this a difficult task. The common practices of overlaying exclusing regions and sampling the parameter space using random or grid samplers are insufficient and much information is lost along the way. Composite likelihoods that keep all the information about experimental data, smart sampling strategies to explore large and intricate parameter spaces~\cite{DarkMachinesHighDimensionalSamplingGroup:2021wkt}, and a rigorous treatment of statistics are critical features that significantly improve and simplify this task~\cite{AbdusSalam:2020rdj}. And for this purpose many global fitting tools have been developed that include those features and that are targeted, though not exclusively, to BSM models. 

Bayesian inference has been a staple in astrophysics for many years, and there are many public tools developed for that purpose, but in the field of BSM phenomenology the tendency has been to develop private fitters for specific purposes, e.g. flavour fits, neutrino oscillations, SUSY, etc. Only in recent years various public inference tools have been released. The first ``wave'' of publicly available global fitters, such as \textsf{Fittino}~\cite{Bechtle:2012zk}, \textsf{SuperBayes}~\cite{RuizdeAustri:2006iwb} or \textsf{EasyScanHEP}~\cite{Han:2016gvr}, were developed to fit various SUSY models to LHC data, and some are still in use today.

The revolution in the field of BSM inference came with the advent of model-independent tools. The front-runner for SM fits is \textsf{HEPFit}~\cite{DeBlas:2019ehy}, built upon a Bayesian Markov chain Monte Carlo, which includes a vast library of electroweak precision, Higgs and flavour observables, and can also work with BSM models~\cite{Chiang:2018cgb,Chowdhury:2017aav}. Nevertheless, the indisputable leader of BSM inference studies is the \textsf{GAMBIT} tool~\cite{GAMBIT:2017yxo,Kvellestad:2019vxm}. \textsf{GAMBIT} contains a huge library of likelihoods and observables for collider physics~\cite{GAMBIT:2017qxg}, dark matter~\cite{GAMBITDarkMatterWorkgroup:2017fax}, flavour~\cite{GAMBITFlavourWorkgroup:2017dbx}, precision observables~\cite{GAMBITModelsWorkgroup:2017ilg}, neutrino physics~\cite{Chrzaszcz:2019inj} and cosmology~\cite{GAMBITCosmologyWorkgroup:2020htv}, as well as an extensive database of models and sampling strategies, both frequentist and Bayesian~\cite{Martinez:2017lzg}. \textsf{GAMBIT} has been employed in studies of a large variety of BSM modes, including SUSY models~\cite{GAMBIT:2017snp,GAMBIT:2017zdo,GAMBIT:2018gjo}, Higgs-portal DM models~\cite{GAMBIT:2017gge,Athron:2018ipf,GAMBIT:2018eea}, EFT DM models~\cite{GAMBIT:2021rlp}, axion and axion-like particle models~\cite{Hoof:2018ieb,Athron:2020maw}, models with right-handed neutrinos~\cite{Chrzaszcz:2019inj}, fits of neutral flavour anomalies~\cite{Bhom:2020lmk} and cosmological fits of neutrino masses~\cite{GAMBITCosmologyWorkgroup:2020rmf}.

\section{Lagrangians and Model Building}

In addition to the software developed for the computation of observables, there is a small collection of tools that go one step further, and attempt to bridge the gap between the full Lagrangian of the BSM theory and the required amplitudes and cross sections. Some of these tools, such as \textsf{FeynRules}~\cite{Christensen:2008py, Alloul:2013bka}, \textsf{SARAH}~\cite{Staub:2008uz,Staub:2012pb,Staub:2013tta} and \textsf{LanHEP}~\cite{Semenov:1996es,Semenov:2002jw,Semenov:2008jy}, compute the Feynman rules directly from the Lagrangian, while others, such as \textsf{FeynArts/FormCalc}~\cite{Mertig:1990an, Kublbeck:1990xc, Hahn:2000kx} or \textsf{CalcHEP}~\cite{Belyaev:2012qa}, use those to compute the required amplitudes. Although widely used, the main disadvantage of most of these tools is that they are written in \textsf{Mathematica}, which is a proprietary software framework and it is notably slower than other interpreted or compiled languages. Very recently a new tool has been released, \textsf{MARTY}~\cite{Uhlrich:2020ltd}, a fully open-source framework that uses symbolic computations in \textsf{C++} to obtain amplitudes and cross sections from the Lagrangian for any BSM theory. Finally, the last step in the BSM toolchain is to automatically connect the amplitudes, cross sections, etc, obtained by the Lagrangian-level tools, with the calculation of likelihoods and observables. This step is most often done by hand, by manually embedding the code or expressions outputted by the above tools where needed. Fortunately, an automatic integration of this step can now be achieved with \textsf{GUM}~\cite{Bloor:2021gtp}. By interfacing with \textsf{SARAH} and \textsf{FeynRules}, \textsf{GUM} extracts all information about particles, parameters and expressions, auto-generates code in \textsf{GAMBIT}, and uses it to impose DM, collider and vacuum stability constraints. Using both \textsf{GUM} and \textsf{GAMBIT} one can start with a Lagrangian and end with a complete likelihood analysis without writing a single line of code.

\section{Summary}

In this conference article I have introduced a selection of tools for BSM physics and described their most recent developments. The state of the communication between the theory and experimental communities is currently at its best, and that improves the quality and relevance of the results from recasting tools. There is a community trend towards the development of model-independent tools, as it is the case for some of the DM comprehensive tools, as well as the leading global fitting frameworks, \textsf{HEPFit} and \textsf{GAMBIT}. And lastly, with \textsf{MARTY} and \textsf{GUM} we have managed to complete the BSM toolchain, from Lagrangian to likelihoods, using only open-source software.

\bibliographystyle{JHEP}
\bibliography{LHCPGonzalo}

\end{document}